\documentstyle[12pt]{article}
\begin{document}
\baselineskip 8mm

\title {\bf Shannon information, LMC complexity and R\'enyi entropies: a straightforward approach}
\author{Ricardo L\'{o}pez-Ruiz$^{(1,2)}$ \\
                                   \\
(1) Department of Computer Sciences, \\
   Facultad de Ciencias - Edificio B \\
(2) Instituto de Biocomputaci\'on y F\'{\i}sica de Sistemas Complejos (BIFI), \\
  Universidad de Zaragoza, 50009-Zaragoza, Spain.}
\date{ }
\maketitle

\begin{center} {\bf Abstract} \end{center}

The LMC complexity, an indicator of {\it complexity} 
based on a probabilistic description, is revisited. 
A straightforward approach allows us to establish the time evolution of
this indicator in a near-equilibrium situation and 
gives us a new insight for interpreting the LMC complexity
for a general non equilibrium system. 
Its relationship with the R\'enyi entropies is also explained.
One of the advantages of this indicator is that its calculation does not require a 
considerable computational effort in many cases of physical and biological interest.

{\small {\bf Electronic mail:} rilopez@unizar.es}

\newpage
\section{Shannon information}

\indent Entropy plays a crucial theoretical role in physics of macroscopic equilibrium
systems. The probability distribution of accessible states of a constrained
system in equilibrium can be found
by the inference principle of maximum entropy \cite{jaynes}.
The macroscopic magnitudes and the laws that relate them can be calculated
with this probability distribution by standard statistical mechanics techniques.

The same scheme could be thought for extended systems far from equilibrium,
but in this case we do not have neither a method to find the
probability distribution nor the knowledge of the relevant magnitudes 
bringing the information that can predict the system's behavior.
It is not the case, for instance, with the metric properties 
of low dimensional chaotic systems by means of the Lyapunov exponents, 
invariant measures and fractal dimensions \cite{badii}.

Shannon information or entropy $H$ \cite{shannon} can still be used as 
a magnitude in a general situation with $N$ accessible states:
\begin{equation}
H=-K\sum_{i=1}^{N}p_i\log{p_i}
\label{eq0}
\end{equation}
with $K$ a positive real constant and $p_i$ the normalized 
associated probabilities, $\sum_{i=1}^{N}p_i=1$.
An isolated system in equilibrium presents equiprobability, 
$p_i=1/N$ for all $i$, among its accessible states and 
this is the situation of maximal entropy,
\begin{equation}
H_{max}=K\log N.
\end{equation}
If the system is out of equilibrium, the entropy $H$ can be expanded
around this maximum $H_{max}$:
\begin{equation}
H(p_1,p_2,\ldots,p_N)=K\log N - \frac{NK}{2}\sum_{i=1}^{N}
\left(p_i-\frac{1}{N}\right)^2 +
\ldots = H_{max} - \frac{NK}{2} D + \ldots
\label{eq1}
\end{equation}
where the quantity $D=\sum_i (p_i-1/N)^2$, that we call {\it disequilibrium},
is a kind of distance from the actual system configuration to the equilibrium.
If the expression (\ref{eq1}) is multiplied by $H$ we obtain:
\begin{equation}
H^2 = H\cdot H_{max} - \frac{NK}{2}\; H\cdot D + K^2 f(N,p_i),
\end{equation}
where $f(N,p_i)$ is the entropy multiplied by the rest of the Taylor expansion 
terms, which present the form ${1\over N}\sum_i (Np_i-1)^m$ with $m>2$. 
If we rename $C=H\cdot D$,
\begin{equation}
C = cte\cdot H\cdot (H_{max} - H) +  K \bar{f}(N,p_i),
\label{eq5}
\end{equation}
with $cte^{-1}=NK/2$ and $\bar{f}=2f/N$. The idea of distance for 
the disequilibrium is now clearer if we see that  
$D$ is just the real distance $D\sim (H_{max}-H)$ for systems 
in the vicinity of the equiprobability.
In an ideal gas we have  $H\sim H_{max}$ and $D\sim 0$, then $C\sim 0$.
Contrarily, in a crystal $H\sim 0$ and $D\sim 1$, but also $C\sim 0$.
These two systems are considered as classical examples of simple models
and are extrema in a scale of disorder ($H$) or disequilibrium ($D$) but
those should present null complexity in a hypothetic measure of
{\it complexity}. This last asymptotic behavior is verified by the variable
$C$ (Fig. 1) and $C$ has been proposed as a  such magnitude \cite{lopez}.
We formalize this simple idea recalling the recent definition 
of {\it LMC complexity} in the next section.

Let us see another important property arising from relation (\ref{eq5}).
If we take the time derivative of $C$ in a neighborhood of equilibrium
by approaching $C\sim H(H_{max}-H)$, then we have
\begin{equation}
{dC\over dt} \sim -2H_{max}{dH\over dt}.
\end{equation}
The irreversibility property of $H$ implies that ${dH\over dt}\geq 0$,
the equality occurring only for the equipartition,
therefore 
\begin{equation}
{dC\over dt} \leq 0.
\end{equation}
Hence, in the vicinity of $H_{max}$, LMC complexity is always decreasing on the evolution
path towards equilibrium, independently of the kind of transition and of the system 
under study. This does not forbid that complexity can increase when the system is very far 
from equilibrium. In fact this is the case in a  general situation as it can be seen,
for instance, in the system presented in Ref \cite{calbet}.

\section{LMC complexity}

Let us assume that at the {\it scale} of observation a system  has N
accessible states $\{x_1,x_2,...,x_N\}$ (N-system) and a probability
distribution $\{p_1,p_2,...,p_N\}$ of each state ($p_i\neq 0$ for all $i$).
Then, at this level of description the knowledge of the underlying
physical laws is "expressed" by a probability distribution
among the accesible states. Shannon \cite{shannon} demonstrated that the
only function that gives the information of a system under the most
elementary assumptions is $H = -K \sum_{i=1}^{N} p_i\log p_i$.
It is easy to find out that the {\it information}
$H$ contained in a crystal  is $H_c \sim 0$, while for an isolated gas
$p_i\sim 1/N$ and then $H_g\sim K\log N$ that represents the maximum of
information for a system with $N$ states. Any other N-system will have
information between those two extrema. \par
The {\it disequilibrium} $D$ of a system can be taken as some kind of
distance to an equiprobable distribution. Two conditions are required to this
magnitude $D>0$  (positive measure of complexity) and $D=0$ in the limit
of equiprobability. The easier solution is to add the quadratic distances of
each state to the equiprobability, i.e., $D = \sum_{i=1}^{N}(p_i -
\frac{1}{N})^2$. This function will be a maximum for a crystal and zero (by
construction) for an ideal gas. Any other N-system will have
disequilibrium between these two extrema. \par
Following the discussion in the introduction the definition of {\it
LMC complexity} ($C$) is presented \cite{lopez}:
\begin{equation}
C = H \cdot D = -\left ( K\sum_{i=1}^{N} p_i\log p_i \right ) \cdot
\left (\sum_{i=1}^{N}(p_i - \frac{1}{N})^2 \right )
\end{equation}
This definition fits the intuitive arguments and gives $C\sim 0$ for the two
systems (a perfect crystal and the ideal gas) discussed herein. Any other
system will have an intermediate behavior and therefore $C>0$. At {\it
different scales} a different number of states (a different probability
distribution) are accesible to the system and therefore different $H$ and $D$.
Therefore, the magnitude complexity defined is {\it scale-dependent} as
expected.

Direct simulations of the definition
of $C$ and its comparison with $H$ for several systems in different 
contexts has been presented in Ref. \cite{lopez,calbet}.
For a 2-system an analytical expression for the curve $C (H)$ is obtained. 
For $N>2$ the relationship between $H$ and $C$ is not
univoque anymore. Many different distributions $\{p_i\}$ can have associated
the same information $H$ but different complexity $C$.  
But a similar form as the one found in the 2-system case is
recovered for a N-system when 
the maximum complexity $C_{max}(H)$ is calculated for each $H$. 
These curves, represented in a
normalized way $\bar{C}_{max}(\bar{H})$, with normalization
$K=1/\log N$, have been numerically computed 
for the cases $N=2,3,5,7$ in Ref. \cite{lopez}.
When $N$ tends towards infinity \cite{calbet} the maximum disequilibrium 
scales as $(1-\bar{H})^2$ and the maximum complexity 
tends to
\begin{equation}
\label{eq:cmaxlim}
\lim_{N \rightarrow \infty} \bar{C}_{max} = 
\bar{H} \cdot (1-\bar{H})^2.
\end{equation}
The limit of the minimum disequilibrium and complexity
vanishes,
\begin{equation}
\label{eq:cminlim}
\lim_{N \rightarrow \infty} \bar{C}_{min} = 0.
\end{equation}
In general, in the limit $N \rightarrow \infty$,
the complexity is not a trivial function of the entropy,
in the sense that for a given $H$ there exists
a range of complexities between $0$ and $C_{\rm max}(H))$
(Eqs. (\ref{eq:cmaxlim}-\ref{eq:cminlim})).
In particular, in this asymptotic limit,
the maximum of $\bar{C}_{max}$ is found when
$\bar{H}=1/3$, which gives a maximum of the maximum complexity
of $\bar{C}_{max}=4/27$ and confirms 
the numerical calculation of Ref. \cite{anteneodo}.
This value is reached when the distribution presents a dominant state with 
probability $p_{max}={2\over 3}$ and the rest of the infinitely many states is 
a uniform 'sea' of equal probability. We say that this distribution
is a type-like 'Rey-Pueblo' configuration.

An attempt of extending the LMC complexity for continuous systems
has been performed in Ref. \cite{catalan}. When the number of states available for
a system is a continuum then the natural representation is a continuous distribution.
In this case, the entropy can become negative.
The positivity of $C$ for every distribution is recovered by taking the exponential
of $H$. If we define $\hat{C}=\hat{H}\cdot D=e^{H}\cdot D$ as an extension of $C$
to the continuous case interesting properties characterizing the indicator $\hat{C}$ appear.
Namely, its invariance under translations, rescaling transformations and replication
convert $\hat{C}$ in a good candidate to be considered as an indicator bringing essential
information about the statistical properties of a continuous system.

The most important point is that the definition should work in
systems out of equilibrium. We present a significative example \cite{lopez} for the
logistic map, the typical chaotic system in which the transition from chaos to
a 3-period orbit via intermittency is know to be extremely "complex". This is
due to the fact that the intermittent bursts are more and more improbable and
impredictible when the transition point is approached. After the transition
point a period three orbit stabilizes and the dynamics becomes simple.
(Complexity is calculated by means of the binary sequences issued from the
numerical simulations of the map, using the natural partition). We see in 
Fig. 2 that the values of C in the intermittency transition point recalls 
a second order phase transition.

\section{R\'enyi entropies}

Generalized entropies have been introduced by R\'enyi \cite{renyi} in the form of
\begin{equation}
I_q={1\over 1-q}\log\left(\sum_{i=1}^N p_i^q\right),
\end{equation}
where $q$ is an index running over  all the integer values. By differentiating $I_q$ with 
respect to $q$ a negative quantity is obtained independently of $q$, then
$I_q$ monotonously decreases when $q$ increases. 

The R\'enyi entropies are an extension of the Shannon information $H$.
In fact, $H$ is obtained in the limit $q\rightarrow 1$:
\begin{equation}
H=I_1=\lim_{q\rightarrow 1} I_q=-\sum_{i=1}^{N}p_i\log{p_i},
\end{equation}
where the constant $K$ of Eq. (\ref{eq0}) is considered to be the unity.
The disequilibrium $D$ is also related with $I_2=-\log\left(\sum_{i=1}^N p_i^2\right)$. 
We have that 
\begin{equation}
D=\sum_{i=1}^N p_i^2-{1\over N}=e^{-I_2}-{1\over N},
\end{equation}
then the LMC complexity is
\begin{equation}
C=H\cdot D=I_1\cdot\left(e^{-I_2}-{1\over N}\right).
\end{equation}
The behavior of $C$ in the neighborhood of $H_{max}$ takes the form 
\begin{equation}
C\sim {1\over N}(\log^2N-I_1I_2),
\end{equation}
The obvious generalization of the R\'enyi entropies for a normalized continuous
distribution $p(x)$ is
\begin{equation}
I_q={1\over 1-q}\log\int [p(x)]^q dx.
\end{equation}
Hence, 
\begin{eqnarray}
H & = & I_1=-\int p(x)\log p(x) dx, \\
D & = & e^{-I_2}=\int [p(x)]^2 dx.
\end{eqnarray}
The dependence of $\hat{C}=e^H\cdot D$ 
with $I_1$ and $I_2$ yields
\begin{equation}
\log\hat{C}= (I_1-I_2).
\end{equation}
This indicates that a family of different indicators could derive from
the differences established among R\'enyi entropies with different $q$-indices.

The invariance of $\hat{C}$ under rescaling transformations implies
that this magnitude is conserved in many different processes.
For instance, the initial Gaussian-like distribution will continue to be 
Gaussian in a classical diffusion process. Then $\hat{C}$ is constant 
in time: ${d\hat{C}\over dt} = 0$, and we have:
\begin{equation}
{dI_1\over dt} = {dI_2\over dt}.
\end{equation} 
The equal losing rate of $I_1$ and $I_2$ is the cost to be paid in order
to maintain the shape of the distribution associated to the system and, hence, 
all its statistical properties will remain unchanged during its time evolution.

\section{Discussion and conclusions}

A definition of complexity (LMC complexity) based on a probabilistic description
of physical systems has been revisited. This definition contains basically an
interplay between the {\it information} contained in the system and the
{\it distance to equipartition} of the probability distribution representing
the system. Besides giving the main features of an intuitive notion of
complexity, we showed that it allows to successfully discern situations
considered as complex in systems of a very general interest.
In particular it proved to be useful for quantifying the complexity in some
statistical and laser systems \cite{lopez1,martin}.
Its relationship with the Shannon information and the generalized R\'enyi entropies 
has been shown to be explicit. Moreover it has been possible to establish the
decrease of this magnitude when a general system evolves from a near-equilibrium 
situation to the equipartition. 

Many different notions of complexity have been proposed until now, mainly
in the context of computational and social sciences. Most of these
definitions present either operational difficulties or arise logical problems. 
The main advantage of LMC complexity is its generality and the fact that 
it is operationally simple and do not
require a big amount of calculations. This advantage has been
worked out in different examples, such as the study of the time evolution 
of $C$ for a simplified model of an isolated gas, the "tetrahedral gas" \cite{calbet},
the slight modification of $C$ as an effective method by which the
complexity in hydrological systems can be identified \cite{guozhang}, 
the attempt of generalize $C$ in a family of simple complexity 
measures \cite{shiner}, some statistical features
of the behavior of $C$ for DNA sequences \cite{zuguo} and
some wavelet-based informational tools used to analyze  the 
brain electrical activity in epilectic episodes in the plane of 
coordinates $(H,C)$ \cite{rosso}.
We are convinced that it provides a
useful way of thinking and it can help in the future to gain more insight
on the physical grounds of models with potential biological interest.

{\bf Acknowledgements}

The author thanks X. Calbet, H.L. Mancini, J.L. L\'opez, R.G. Catal\'{a}n,
J. Garay and A.F. Pacheco for very fruitful discussions on different aspects
of this subject of complex systems.

%\newpage
\begin{center} {\bf Figure Captions} \end{center}

{\bf 1.} Sketch of physical intuition of the magnitudes "information" (H)
and "disequilibrium" (D) between the two simple systems: crystal and
ideal gas. Also, intuitive behavior required for the magnitude {\it
complexity}.
The quantity $C=H\cdot D$ was proposed as such a magnitude.

{\bf 2.} Behavior of the complexity $C$ in the transition point
($p_t\sim 3.8284$) where the system (logistic map) goes from a chaotic
dynamics by intermittency ($p<p_t$) to a period three orbit ($p>p_t$).

\end{document}